\documentstyle[aps,pre,epsf,eqsecnum]{revtex}
\begin{document}
\draft
\title
{
Solitons and diffusive modes in the noiseless Burgers equation\\
Stability analysis
}
\author{Hans C. Fogedby}
\address{
\thanks{Permanent address}
Institute of Physics and Astronomy,
University of Aarhus, DK-8000, Aarhus C, Denmark\\
and\\
NORDITA, Blegdamsvej 17, DK-2100, Copenhagen {\O}, Denmark
}
\date{\today}
\maketitle
\begin{abstract}
The noiseless Burgers equation in one spatial dimension is analyzed
from the point of view of a diffusive evolution
equation in terms of  nonlinear soliton  modes and linear diffusive modes.
The transient evolution of the profile is interpreted as a gas
of {\em right hand} solitons connected by ramp solutions with superposed
linear diffusive modes. This picture is supported by a
linear stability analysis of the soliton mode. The
spectrum and phase shift of the diffusive modes
are determined. In the presence of the soliton the diffusive
modes develop a gap in the spectrum and are phase-shifted
in accordance with Levinson's theorem. The spectrum also
exhibits a zero-frequency translation or Goldstone mode
associated with the broken translational symmetry.
\end{abstract}
\section{Introduction}
This is the first of a series of papers dedicated to an analysis
of the Burgers equation in one spatial dimension. 
To set the stage for our main objective, namely
the analysis of the {\em noisy} Burgers equation for the stochastic growth 
of an interface driven by annealed white noise to be presented in the
following paper, we have found it convenient to begin with a brief analysis
of the {\em deterministic} or {\em noiseless} Burgers equation. The reason is
that many of the features of the noiseless case such a nonlinear
soliton modes
etc. also play a decisive role in the noisy case.
The noiseless Burgers equation has the form
\begin{equation}
\frac{\partial u}{\partial t} = \nu\nabla^2 u + \lambda u \nabla u~ .
\end{equation}
Here $\nu$ is a damping constant controlling the strength of the linear
dissipative term. The parameter $\lambda$ characterizes the strength
of the nonlinear mode coupling term.

For $\lambda = -1$ the equation (1.1) was originally proposed by
Burgers \cite{1} as a model for irrotational or vorticity-free
hydrodynamics in order to describe one dimensional turbulence \cite{2,3,4}.
In the present context for general $\lambda$
we consider Eq. (1.1) as providing a description of the slope
field $u=\nabla h$ for a growing interface governed by the noiseless
or deterministic Kardar-Parisi-Zhang equation \cite{5}
\begin{equation}
\frac{\partial h}{\partial t} = \nu\nabla^2 h + \frac{\lambda}{2}(\nabla h)^2~ .
\end{equation}
The Burgers equation (1.1) has been the subject of numerous studies as
an initial value problem, e.g., ref. \cite{6}, and is still of 
considerable current interest \cite{7}.

It is an interesting property
of the Burgers equation (1.1) that it
is exactly soluble in the sense that the nonlinear Cole - Hopf
transformation \cite{8,9}
\begin{eqnarray}
w =&& \exp{\left[\frac{\lambda}{2\nu}\int^x u dx\right]}\\
u =&& \frac{2\nu}{\lambda}\nabla\log{w}
\end{eqnarray}
allows for an exact mapping onto the linear diffusion equation
\begin{equation}
\frac{\partial w}{\partial t} = \nu\nabla^2 w 
\end{equation}
which can be simply analyzed. For given initial data
$u(x,t=0) = u_0(x)$ we thus have
\begin{equation}
w(x,t) = 
\int G(x-x',t)\exp{\left[\frac{\lambda}{2\nu}\int^{x'}u_0dx''\right]} ~ ,
\end{equation}
where $G(x,t)$ is the Green's function for the diffusion equation
(1.5)
\begin{equation}
G(x,t) = [4\pi\nu t]^{-1/2}\exp{\left[-\frac{x^2}{4\nu t}\right]}
\end{equation}
and $u$ is given by (1.4).

The Cole - Hopf transformation (1.3-4) thus permits a rather 
complete analysis of the Burgers equation. The relaxational dynamics
of the equation is controlled by solitons connected by 
smooth regions;
in the inviscid limit $\nu\rightarrow 0$ the solitons become shocks
connected by ramps \cite{2,4,5,6}. 
Although the nonlinear character of the equation prevents the application
of a superposition principle we can still from a qualitative point of
view envisage that
an initial
configuration $u_0(x)$ ``contains'' a certain number of solitons or
smoothed shocks. As time progresses the configuration passes through a 
transient regime dominated by a gas of propagating and coalescing
solitons with superposed linear diffusive modes. At infinite times
the configuration eventually decays owing to the inviscid term in Eq. (1.1).
This qualitative behavior is depicted in Fig. 1.

It is instructive to compare the nonlinear irreversible and dissipative 
Burgers equation (1.1) with the nonlinear reversible and dispersive
evolution equations \cite{10}:

The equation of motion for the  $u^4$ field theory
\begin{equation}
\frac{\partial^2u}{\partial t^2} = \nabla^2 u + m^2 u - \lambda u^3 ~ ,
\end{equation}
the complex nonlinear Schr\"{o}dinger equation
\begin{equation}
i\frac{\partial u}{\partial t} = \nabla^2 u + \lambda |u|^2 u ~ ,
\end{equation}
and the sine-Gordon equation,
\begin{equation}
\frac{\partial^2u}{\partial t^2} = \nabla^2 u + \lambda \sin{u} ~ .
\end{equation}
In addition to the linear dispersive modes obtained for $\lambda =0$ the
above evolution equations all support soliton solutions due to the 
dynamical balance between the linear dispersive term and the 
nonlinear term controlled by $\lambda$. An initial configuration
thus again breaks up into a gas of moving solitons and linear modes. 
In the case of the sine-Gordon and nonlinear Schr\"{o}dinger
equations, the solitons preserve their identity under collisions 
owing to 
the complete integrability of these systems; this is not the case for 
the $u^4$ field equation, here the solitons become deformed under collisions.

In the present paper we analyze the Burgers equation (1.1) from the 
point of view of a soliton-carrying dissipative evolution equation. 
The paper is organized in the following way. In section II we summarize 
the general properties of the Burgers equation.
In section III we discuss in particular the soliton 
solution and comment on
the morphology of a growing interface. In section IV we present a linear
stability analysis of the Burgers equation and discuss the translation
mode and the diffusive scattering modes. In section V we summarize our
results and present a conclusion.
\section{General properties}
The Burgers equation (1.1) has the form of a nonlinear diffusive 
evolution equation with a linear diffusive term controlled by the damping
or viscosity $\nu$ and a nonlinear mode coupling term characterized
by $\lambda$. In the context of fluid motion the nonlinear term
gives rise to convection as in the Navier Stokes equation \cite{1,2,3}; 
for an
interface the term corresponds to a slope dependent growth \cite{11}.

Under time reversal $t\rightarrow -t$ and the transformation 
$u\rightarrow -u$ the
equation is invariant provided $\nu\rightarrow -\nu$. This indicates
that the linear diffusive and the nonlinear convective or
growth terms play a completely different role. The diffusive term 
is intrinsically 
irreversible whereas the growth term, corresponding to a mode coupling,
gives rise to a cascade in wavenumber space and a genuine transient
growth. The transformation 
$t\rightarrow -t$ is absorbed in $u\rightarrow -u$ or, alternatively, 
$\lambda\rightarrow -\lambda$, corresponding to a change of growth direction.

We also note that the equation is invariant under the parity
transformation $x\rightarrow -x$ provided $u\rightarrow -u$. 
This feature is related 
to the presence of a single spatial derivative in the growth term
and implies that the equation only supports solitons or shocks with one
parity, that is {\em right hand} solitons. We mention in passing that parity
invariance is restored in the case of the noisy Burgers
equation. This interesting aspect will be considered in the next paper.

The Burgers equation is also invariant under a more fundamental symmetry, 
namely the Galilean symmetry group. In fact, boosting the equation to
a moving frame with velocity $\lambda u_0$, $x\rightarrow x-\lambda u_0t$
it is easily seen by inspection,
using $\partial/\partial t\rightarrow\partial/\partial t +\lambda u_0\nabla$, 
that the equation remains invariant provided we shift the amplitude 
$u$ by  $u_0$, i.e., $u\rightarrow u+u_0$. We note that unlike the $u^4$ and
Sine-Gordon equations (1.8) and (1.10) which
are invariant under a linear Lorentz transformation with no change 
in the field amplitude and the nonlinear Schr\"{o}dinger equation 
(1.9)
which is invariant under a Galilean transformation accompanied by
a space and time dependent phase shift in the wave function, the Galilean
transformation of the Burgers equation is nonlinear in the sense that
the slope field is also shifted. Furthermore, the nonlinear coupling 
strength $\lambda$ enters 
explicitly in the Galilean symmetry group.

In the absence of the nonlinear growth term for $\lambda =0$ the 
Burgers equation (1.1) reduces to the linear diffusion equation (1.5)
supporting linear diffusive modes $u\propto\exp{(-i\omega t\pm ikx)}$
with quadratic dispersion
\begin{equation}
\omega = -i\nu k^2~ .
\end{equation}
An initial plane wave configuration thus decays with an envelope
$\exp{(-\nu k^2 t)}$.
More explicitly, defining the Laplace-Fourier transform
\begin{equation}
u(k,\omega) = \int dxdt\exp{(i\omega t -ikx)} u(x,t)\eta(t) ~ ,
\end{equation}
where $\eta(t)$ is the step function, i.e., $\eta(t)=1$ for $t>0, 
\eta(t)=0$ for $t<0$, and $\eta(0) = 1/2$,
and denoting the initial slope configuration by $u_0(k)=u(k,t=0)$, we have the
solution
\begin{equation}
u(k,\omega) = \frac{u_0(k)}{-i\omega + \nu k^2}
\end{equation}
displaying a diffusive pole given by Eq. (2.1). For the temporal correlations
we obtain in particular
\begin{equation}
\langle u(k,t)u(-k,t')\rangle = 
\langle u_0(k)u_0(-k)\rangle\exp{[-(t+t')\nu k^2]}
\end{equation}
where $\langle\cdots\rangle$ denotes an average over 
the distribution of 
initial values.

On the other hand, in the inviscid limit for vanishing damping 
$\nu\rightarrow 0$
the Burgers equation (1.1) takes the form
\begin{equation}
\frac{\partial u}{\partial t} = \lambda u\nabla u
\end{equation}
which has an exact solution given by the implicit equation
$u = F(x+\lambda u t)$, where $F$ is an arbitrary profile. Since the
propagation velocity $\lambda u$ thus increases with the amplitude $u$
it follows
that an
initial configuration $u_0 =F(x)$  breaks and that shock waves
are generated. From the form of the exact solutions it also follows that 
the shocks develop with right parity, i.e., a positive discontinuity
in $u$. As mentioned above this is consistent with the  parity
breaking properties of
Eqs. (1.1) and (2.5). Searching for a static solution of the
form $u = A+B\eta(x-x_0)$ we find by insertion 
\begin{equation}
u(x) = |u_+|\eta(x-x_0)
\end{equation}
with arbitrary amplitude $|u_+|$; a  moving shock is then  obtained by
applying a Galilean boost, i.e., $x\rightarrow x-\lambda u_0 t$, 
$u\rightarrow u +u_0$, yielding shock solutions of Eq. (2.5).
It is also easily seen that Eq. (2.5) supports ramp solutions of the
form
\begin{equation}
u(x) = \mbox{const}~ - \frac{x}{\lambda t} ~ .
\end{equation}

The general picture that emerges in the inviscid limit 
$\nu\rightarrow 0$ is thus that
an initial configuration $u_0 = F(x)$  breaks up into a series of 
{\em right hand} shocks 
connected by ramps; this picture is in fact substantiated
by a steepest descent analysis of the Cole-Hopf transformation (1.3-4) in
the inviscid limit \cite{6}. The time evolution is similar to the 
one depicted in
Fig. 1 except that the solitons are sharp. Furthermore, in the absence of
dissipation
the transient regime
extends to infinite times.
In terms of the height field $h = \int u dx$, the morphology consists
of cusps connected by convex parabolic segments \cite{5,11}.

In the presence of damping or viscosity this picture is not radically
changed. The damping leads to an overall relaxation of the initial
configuration 
where the energy, based on the hydrodynamical definition $\int u^2 dx$
(kinetic energy) is mainly dissipated in the shocks; in other words,
the nonlinear mode coupling term gives rise to spatially confined
{\em hot zones} for energy dissipation associated with the solitons.
\section{The soliton solution}
Although the Burgers equation (1.1) admits an 
exact solution by means of the
Cole-Hopf transformation (1.3-4), we find it useful for our
purposes to  approach Eq. (1.1) as a diffusive
nonlinear evolution equation and following the corresponding analysis
of Eqs. (1.8-10) search for permanent profile soliton 
solutions \cite{10}. 
Setting $u(x,t) = u(x-vt)$ where $v$ is the propagation velocity
and using $\partial/\partial t = -v\nabla$ Eq. (1.1) can be integrated
once
\begin{equation}
-vu = \nu\nabla u + \frac{\lambda}{2}u^2 + \mbox{const} ~ .
\end{equation}
Furthermore,  imposing the boundary conditions 
$u=u_\pm,\nabla u\rightarrow 0$ for 
$x\rightarrow\pm\infty$,
appropriate for a single soliton solution, and subtracting Eq. (3.1)
for $x\rightarrow\pm\infty$,  we obtain the soliton condition
\begin{equation}
u_+ + u_- = -\frac{2v}{\lambda}
\end{equation}
relating the propagation velocity $v$ of the soliton to 
the boundary values $u_\pm$.
We note that Eq. (3.2) is consistent with the fundamental nonlinear
Galilean symmetry, being
invariant under the transformation 
$v\rightarrow v+\lambda u_0, u_\pm\rightarrow u_\pm -u_0$. 
In terms of the boundary 
values $u_\pm$ we can also express Eq. (3.1) in the form
\begin{equation}
\nabla u =\frac{\lambda}{2\nu}(u_+-u)(u-u_-)
\end{equation}
which implies a positive slope of $u$ between the boundary values 
$u_\pm$ and $u_+>u_-$, corresponding to a {\em right hand} or positive 
parity soliton in accordance with the symmetry property discussed
in section II. 
In the static limit $v = 0$
Eq. (3.2) implies $u_+ = -u_-$, that is a symmetric soliton and 
by quadrature
Eq. (3.3) yields
the static soliton solution
\begin{eqnarray}
u_0(x) =&& u_+\tanh{[k_s(x-x_0)]}\\
k_s =&& \lambda u_+/2\nu ~.
\end{eqnarray}
We have introduced the characteristic wavenumber $k_s$ setting the inverse
length scale of the static soliton; $x_0$ denotes the center of mass
position. The width of the soliton is of order $1/k_s$ and,  unlike the
$u^4$ or sine-Gordon soliton related to the amplitude $u_+$. In the inviscid
limit $\nu\rightarrow 0$, the wave number $k_s\rightarrow\infty$ and 
the soliton approaches the
shock discontinuity given by  Eq. (2.6).

A moving soliton is obtained by applying a Galilean boost 
$x\rightarrow x-\lambda u_0 t$ and shifting the profile by $u_0$. In terms of
the boundary values $u_\pm$ we thus obtain the propagating soliton
solution
\begin{equation}
u_0(x,t) = \frac{u_++u_-}{2} + \frac{u_+-u_-}{2}
\tanh{\left[\frac{\lambda}{4\nu}(u_+-u_-)(x-vt-x_0)\right]}
\end{equation}
with  velocity $v$ given by Eq. (3.2). In Fig. 2 we have depicted a 
single soliton
solution of the Burgers equation and the associated smoothed cusp profile for
the height field of a growing interface.
\section{Linear stability analysis}
In order to investigate the properties of the linear diffusive modes in the
presence of the nonlinear soliton mode 
we here perform a
linear stability analysis.  Since a Galilean transformation
allows for a boost to a finite propagation velocity it is sufficient 
to consider the stability of the static soliton $u_0$ in Eq. (3.4).
Setting $u = u_0 +\delta u$ we obtain by inserting in the Burgers equation
(1.1) to linear order in $\delta u$ the equation of motion for the fluctuations
about the soliton profile
\begin{equation}
\frac{\partial\delta u}{\partial t} = \nu\nabla^2\delta u
+
\lambda u_0\nabla\delta u + \lambda (\nabla u_0)\delta u ~.
\end{equation}
Absorbing the first order derivative term by means of the transformation
\cite{12}
\begin{equation}
\delta u = \frac{\delta \tilde {u}}{\cosh{[k_s(x-x_0)]}}
\end{equation}
and searching for solutions with time dependence 
$\delta u\propto\exp{(-i\omega t)}$ we arrive at the linear eigenvalue problem
\begin{equation}
-i\omega\delta\tilde{u} = \nu\nabla^2\delta\tilde{u}
-
\nu k^2_s\left[1-\frac{2}{\cosh^2{[k_s(x-x_0)]}}\right]\delta\tilde{u} ~.
\end{equation}
This equation has the same form as the one encountered in the linear stability
analysis of the sine-Gordon soliton \cite{10,13}. Interpreted as a
stationary Schr\"{o}dinger equation Eq. (4.3) describes a particle with energy 
$i\omega$ and mass $1/2\nu$ in the exactly soluble Bargman potential
$-2/\cosh^2{x}$. The spectrum is well-known and consists of a single
bound state for $\omega =0$ and a band of scattering states for
$\omega = -i\nu (k^2+k^2_s)$ \cite{13,14}. In Fig. 3 we have sketched 
the potential
and the band of scattering states.
\subsection{The translation mode}
The bound state solution for $\omega = 0$ has the form
$\delta\tilde{u}_{BS}\propto 1/\cosh{[k_s(x-x_0)]}$ and using
Eq. (4.2)
\begin{equation}
\delta u_{BS}\propto\frac{1}{\cosh^2{[k_s(x-x_0)]}},~~~~~ \omega = 0~ .
\end{equation}
This zero frequency mode has a particular significance for 
soliton-carrying systems. Since $\delta u_{BS}\propto(du_0/dx)\delta x$
it is seen that the  mode actually corresponds to an 
infinitesimal translation $\delta x$
of the soliton without changing its shape. This mode thus restores 
the broken
translational invariance resulting from the choice of a particular 
center of mass position $x_0$ for the soliton - a so-called translation mode. 
Such Goldstone modes are quite generally associated with broken
symmetries \cite{10,13,15}.
\subsection{The diffusive scattering modes}
In a similar way the band of scattering states has the explicit form
\begin{equation}
\delta u_{SS}\propto
\frac{\exp{(ikx)}}{\cosh{[k_s(x-x_0)]}}
\frac{k+ik_s\tanh{[k_s(x-x_0)]}}{k-ik_s}
\end{equation}
with dispersion
\begin{equation}
\omega = -i\nu(k^2+k^2_s) ~ .
\end{equation}
The modes form a continuum of spatially decaying diffusive states scattering 
off the soliton. We note that the dispersion law (4.6) for the diffusive 
spectrum compared with Eq. (2.1) for the linear case
has developed a gap $\nu k^2_s = \lambda^2u_+^2/4\nu$ depending on the
soliton amplitude.
For $x\rightarrow\pm\infty$ we have 
$\delta u_{SS}\propto\exp{(ikx + i\delta(k))}\exp{(-k_s|x|)}$ and
$\delta u_{SS}\propto\exp{(ikx)}\exp{(-k_s|x|)}$, respectively, 
where the plane wave 
part is phase shifted by the amount
\begin{equation}
\delta(k) = 2\tan^{-1}{(\frac{k_s}{k})} ~ .
\end{equation}
The transmission coefficient is thus given by $t(k) = \exp{(i\delta(k)}$, i.e.,
$|t(k)|^2 = 1$, yielding the reflection coefficient $r(k) = 0$
since $|r(k)|^2 = 1 - |t(k)|^2 = 0$. Consequently, the soliton acts as
a reflectionless transparent potential on the diffusive modes which 
only experience
a phase shift. We also note that the bound state energy
is given by the pole $k=ik_s$ in the S-matrix $(k+ik_s)/(k-ik_s)$; the residue
in Eq. (4.5) yielding the bound state in Eq. (4.4). Furthermore, imposing
periodic boundary conditions in a system of size $N$ we obtain the density
of diffusive modes
\begin{equation}
\rho = \frac{N}{2\pi} + \frac{1}{2\pi}\frac{d\delta(k)}{dk} ~ .
\end{equation}
For the change of density of states owing to the presence of the soliton
, $\Delta\rho = \rho - N/2\pi$, we have, inserting Eq. (4.7),
\begin{equation}
\Delta\rho = -\frac{1}{\pi}\frac{k_s}{k^2 + k_s^2} ~ .
\end{equation}
We note that $\int\rho dk = N-1$ in accordance with Levinson's
theorem, i.e., the band of diffusive modes is depleted by one mode
corresponding to the zero frequency translation mode. In Fig. 4   
we have
depicted the phase shifted-diffusive mode and the translation mode. 
In Fig. 5 we have shown the  diffusive spectrum (4.6)
and the pole structure in the complex wavenumber plane.

Finally, regarding the stability of the soliton we note that the 
linear diffusive mode is damped according to 
$\delta u_{SS}\propto\exp{(-\Gamma(k)t)}$
with a damping constant $\Gamma(k)=\nu (k^2+k^2_s)$. In the long wavelength
limit $k\rightarrow 0$, $\Gamma(k)$ approaches a constant 
$\nu k_s^2 = \lambda^2u_+^2/4\nu$, i.e., the gap in the diffusive 
spectrum. This result is, however, not 
in conflict with the usual argument of a vanishing $\Gamma(k)$ for
$k\rightarrow 0$, characteristic of a hydrodynamical diffusive mode
\cite{15}. The argument generally follows from the conservation of
the local slope or velocity (momentum) implied by the structure of the
Burgers equation (1.1) which can be written as a local conservation
law
\begin{equation}
\frac{\partial u}{\partial t} = -\nabla j
\end{equation}
with a current density
\begin{equation}
j = -\nabla u - \frac{\lambda}{2} u^2 ~ .
\end{equation}
In the long wavelength limit $k\rightarrow 0$ the conservation law (4.10) 
usually
implies that the damping constant $\Gamma(k)\rightarrow 0$, i.e., an infinitely
long lived mode. This follows from
\begin{equation}
\lim_{k\rightarrow 0}\frac{\partial u(k,t)}{\partial t}
=
\frac{d}{dt}\int u(x,t) dx
=-\int\nabla j dx
\end{equation}
which vanishes 
{\em provided} that there are no currents on the boundaries $x=\pm N/2$. 
However, in the presence of a soliton there is an incoming 
current $j(\pm N/2) = -\lambda/2 u^2_+$ and the 
mode decays with a finite life time $1/\nu k_s^2$ in the long wavelength
limit, corresponding to a gap in the spectrum of $i\omega$.
\section{Summary and conclusion}
In the present paper we have conducted yet a study of the well-documented 
noiseless or deterministic Burgers equation regarding it as a 
nonlinear diffusive evolution equation. Although the
nonlinear Cole-Hopf mapping to a linear diffusive equation in principle
allows for a rather complete analysis of the equation, we have found it
useful to emphasize the solitonic aspects, 
drawing on the parallel with other nonlinear equations such as
the $\phi^4$ and the sine-Gordon equations. Note, however, that
unlike the $\phi^4$ or sine-Gordon equations, where the soliton owes its
stability to a balance between the dispersive effect of the linear term,
tending to break up a wave packet construction, and the cascade effect
in wave number space due to the nonlinear term, stabilizing a particular
wave packet form - the soliton, the Burgers equation is intrinsically
dissipative and an initial configuration will eventually decay due to
dissipation {\em unless} energy is fed into the system. In this regard the
Burgers soliton
is a {\em dissipative} structure in that it owes its 
stability to the energy flux fed by the non-vanishing currents entering
at the boundaries. 
The nonlinear term generating an {\em inverse cascade}
in wave number space thus provides the energy transport to the center of the
soliton where the energy is dissipated and the soliton owes its stability to
the interplay between the linear dissipative term and the nonlinear 
mode coupling term. Nevertheless, it is useful to consider the soliton
as the fundamental {\em elementary excitation} in the Burgers equation
determining the nonlinear nonlocal relaxational aspects. As follows
from a steepest descent analysis in the inviscid limit $\nu\rightarrow 0$, an
initial configuration evolves into a gas of propagating solitons
connected by ramps. The present linear stability analysis then shows
that the linear 
modes which for $\lambda =0$
dominate the relaxational by means of diffusion for $\lambda\neq 0$
become
sub-dominant in the sense that they develop a gap in the $i\omega$ spectrum.

In a subsequent paper we consider the Burgers equation driven by spatially
uniform stochastic noise rather that deterministic currents
at the boundaries. We find that the solitonic aspects still determine
the physics and that the soliton becomes a {\em bona fide} elementary
excitation in the underlying field theory.
\acknowledgements
{
Discussions with M. H. Jensen, T. Bohr, M. Howard,
K. B. Lauritsen and A. Svane
are gratefully acknowledged.
}

\begin{figure}[htb]
\centerline
{
\epsfxsize=12cm
\epsfbox{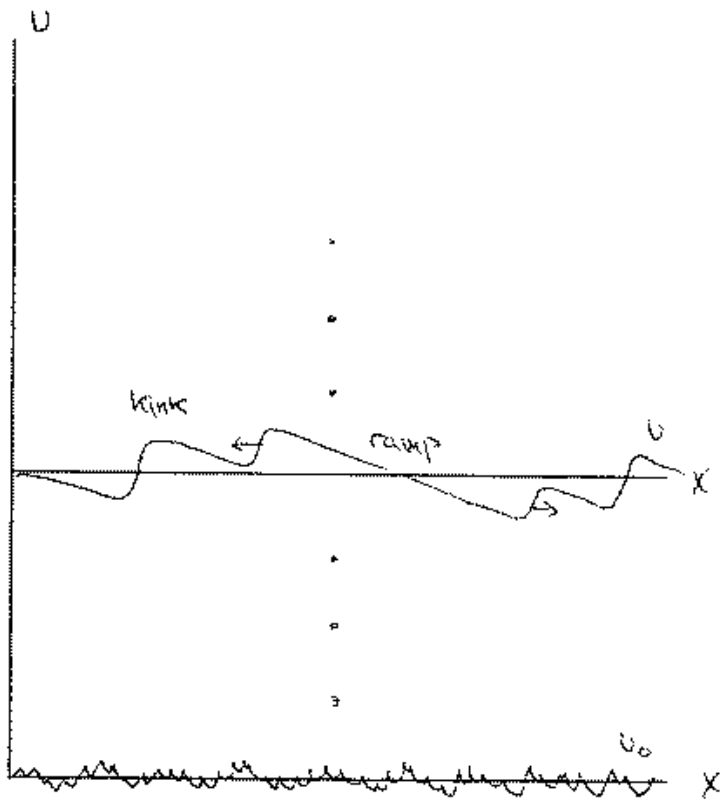}
}
\caption{
Here we show the transient evolution of the slope field $u$ from an
arbitrary initial configuration $u_0$. The transient morphology
consists of propagating {\em right hand} solitons connected by
ramps with superposed
damped diffusive modes. Both the solitons  and diffusive modes transport
energy which is dissipated predominantly at the soliton positions.
At long times the profile decays unless it is driven by currents
at the boundaries, corresponding to non-vanishing slope.
}
\end{figure}
\begin{figure}[htb]
\centerline
{
\epsfxsize=9cm
\epsfbox{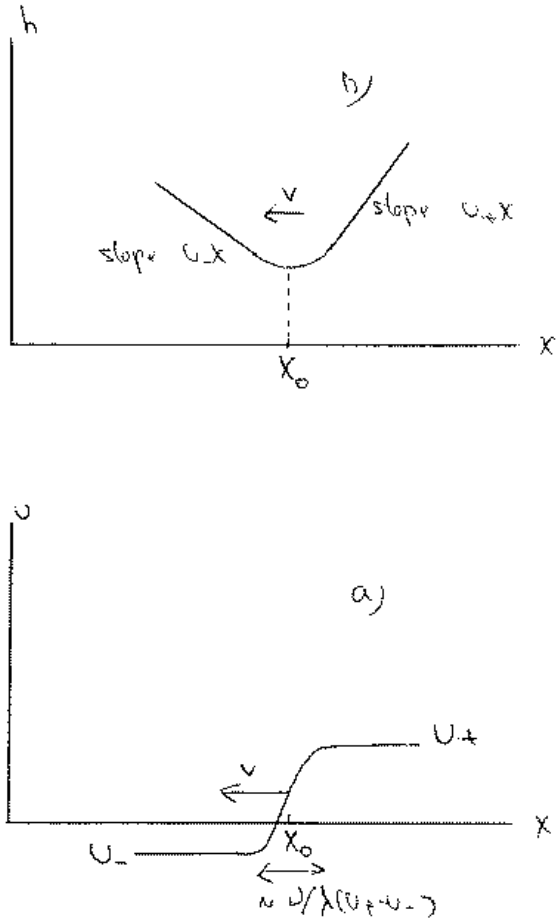}
}
\caption{
In a) we show a single moving soliton profile propagating to the left
and in b) the corresponding smoothed cusp in the growth profile.
This configuration is driven by currents at the boundaries,
corresponding to non-vanishing $u_\pm$ and is persistent in time.
}
\end{figure}
\begin{figure}[htb]
\centerline
{
\epsfxsize=7cm
\epsfbox{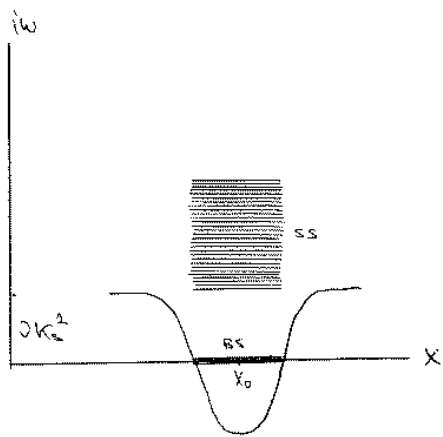}
}
\caption{
We here sketch the reflectionless Bargman potential
$\nu k_s^2[1-2/\cosh^2{[k_s(x-x_0)]}]$. We also show
the associated zero-frequency bound state (BS) or translation mode
and  the band of scattering states above the gap $\nu k_s^2$
in the spectrum.
}
\end{figure}
\begin{figure}[htb]
\centerline
{
\epsfxsize=8cm
\epsfbox{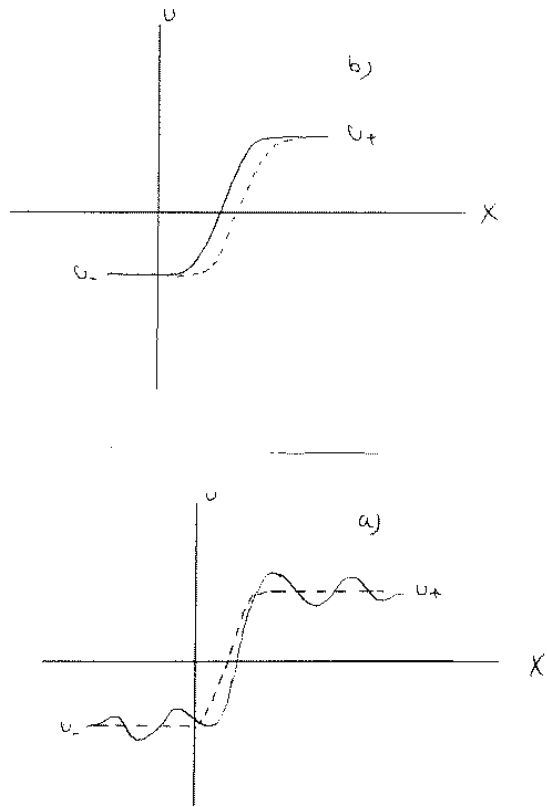}
}
\caption{
In a) we show the phase shifted diffusive mode superposed on
the static  soliton, indicated by a
dashed line. In b) we depict the translation mode giving rise to
a solid displacement of the soliton without changing its shape,
thus lifting the broken translational invariance.
}
\end{figure}
\begin{figure}[htb]
\centerline
{
\epsfxsize=7cm
\epsfbox{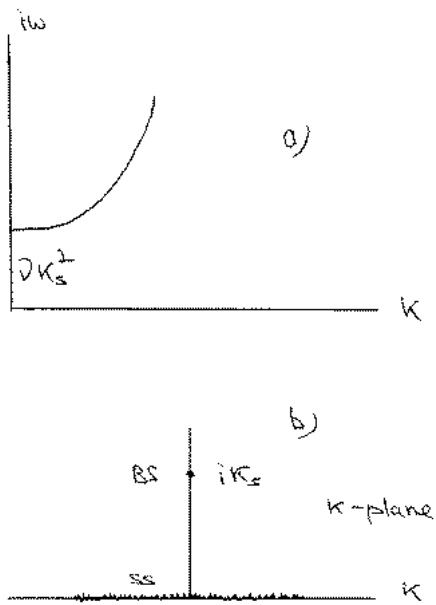}
}
\caption{
In a) we show the spectrum of  diffusive modes in the presence
of a soliton. Unlike the quadratic dispersion for the linear diffusion equation,
the spectrum here exhibits a gap $\nu k_s^2$ depending on the soliton
amplitude. In b) we show the pole structure in the
complex $k$ plane. The real axis corresponds to the band of scattering
states. The pole in the S-matrix on the imaginary axis corresponds to
the bound state or
translation mode.
}
\end{figure}
\end{document}